\documentstyle[12pt]{article}

\def\[{\left\lbrack}
\def\]{\right\rbrack}

\def\({\left(}
\def\){\right)}
\def\ih{\'\i}

\title{A proposal for a generalized canonical osp(1,2) quantization of
dynamical systems with constraints}

\author{Petr M. Lavrov$^a$\thanks{lavrov@tspu.edu.ru} ,
Jorge Ananias Neto$^b$\thanks{jorge@fisica.ufjf.br}
 and Wilson Oliveira$^b$\thanks{wilson@fisica.ufjf.br}
\\ \\
$^a$ Tomsk State Pedagogical University, Tomsk 634041, Russia
\\
$^b$ Departamento de F\ih sica, ICE,
Universidade Federal de Juiz \\de Fora,  36036-330
Juiz de Fora, MG, Brazil}

\date{}

\begin{document}

\maketitle

\begin{abstract}
\noindent The aim of this paper is to consider a possibility of
constructing for arbitrary dynamical systems with first-class
constraints a generalized canonical quantization method based on the
$osp(1,2)$ supersymmetry principle. This proposal can be considered as
a counterpart to the $osp(1,2)$-covariant Lagrangian quantization
method introduced recently by Geyer, Lavrov and M\"ulsch.  The gauge
dependence of Green's functions is studied. It is shown that if the
parameter $m^2$ of the $osp(1,2)$ superalgebra is not equal to zero then
the vacuum functional and $S$ - matrix depend on the gauge. In the limit
$m\rightarrow 0$ the gauge independence of vacuum functional and $S$ -
matrix are restored. The Ward identities related to the $osp(1,2)$
symmetry are derived.

\end{abstract}

\vskip 0.5 cm

\maketitle
\newpage

\setlength{\baselineskip} {20 pt}

\section{Introduction}

The canonical version of extended BRST formalism
\cite{Lavrov1,Lavrov2,Lavrov3} is intended, in principle, to quantize
dynamical systems with constraints. This method is based on a
special type of Hamiltonian that possesses, simultaneously, the so-called
BRST and anti-BRST global symmetries. These requirements can be
implemented by using the global symplectic group $Sp(2)$, in which the
generators of BRST and anti-BRST transformations turn out to form a doublet
of the $Sp(2)$-group.

The study of \cite{Geyer} has proposed the so-called $osp(1,2)$-covariant
Lagrangian quantization. This quantization method is based on a supergroup,
which is larger than the extended BRST supergroup applied in papers
\cite{Lavrov4,Lavrov5,Lavrov6}, and which allows a natural procedure of
including massive terms, needed to circumvent possible infrared
singularities arising in the process of subtracting ultraviolet
divergences, in a manner avoiding the breaking of extended BRST symmetry.

Further development of quantization procedures in the spirit of papers
\cite{Vilk,Bat,Frad} and \cite{Lavrov1,Lavrov2,Lavrov3} makes it possible
to formulate the rules of generalized canonical quantization based on the
global orthosymplectic supergroup $osp(1,2)$. Thus, the present paper should
be considered as a canonical counterpart of the Lagrangian method of
$osp(1,2)$-covariant quantization \cite{Geyer}. We will be concerned in a
possibility of developing a consistent scheme of generalized
canonical quantization, omitting so far all proofs of existence theorems.

In the phase space of arbitrary canonical variables $(P_A,\;Q^A)$ we apply
the usual definition of the Poisson superbracket
\begin{equation}
\{ G,F \} =  {\delta G\over \delta Q^A} {\delta F\over \delta P_A}
- {\delta F\over \delta Q^A} {\delta G\over \delta P_A}
(-1)^{\epsilon(G)\epsilon(F)},
\end{equation}
\vskip .5 cm
\noindent
where $\epsilon(G)$ is the Grassmann parity of a quantity G. The
Grassmann parities of the momenta $P_A$ coincide with those of the
corresponding coordinates $Q^A$: $\epsilon(P_A)= \epsilon(Q^A)= \epsilon_A$.

The Poisson superbracket (1) possesses the following properties:
\begin{equation}
\epsilon(\{ G,F \}) =  \epsilon(G) + \epsilon(F),
\end{equation}
\begin{equation}
\{ G,F \} = - (-1)^{ \epsilon(G) \epsilon(F)}\{ F,G \},
\end{equation}
\begin{equation}
\{\{ G,F \},H\}(-1)^{ \epsilon(G) \epsilon(H)} + cycle (G,F,H)\equiv 0.
\end{equation}
The above relation is the Jacobi identity for the Poisson superbracket.
In Eq.(1) and henceforth, derivatives with respect to the coordinates are
always understood as acting from the right, and those with respect to the
momenta, as acting from the left.

The paper is organized as follows. In Section II we discuss the general
features of canonical $osp(1,2)$ quantization, including the construction of
a unitarizing Hamiltonian invariant under the extended BRST and $Sp(2)$
transformations. Section III is devoted to the question of gauge dependence
and Ward identities. Conclusions are given in Section IV.

\section {Canonical osp(1,2) quantization}

\begin{sloppypar}
Let us consider a dynamical system described by a Hamiltonian
$H_0 =  H_{0}(p_{i},q^{i})$, $\epsilon(H_0) =0$, in the phase space of
initial canonical variables $(p_{i},q^{i})$, $i =  1,2,\dots,n$,
$\epsilon(p_i)= \epsilon_i$ as well as by a set of first-class constraints
$T_{\alpha_0}=T_{\alpha_0}(p_{i},q^{i})$, $\alpha_0=1,2,\dots,m<2n$, with
Grassmann parities $\epsilon(T_{\alpha_0})=\epsilon_{\alpha_0}$.
The following involution relations hold true:
\end{sloppypar}

\begin{eqnarray}
\bigl\{T_{\alpha_0},\,T_{\beta_0}\bigr\} &= & T_{\gamma_0}
U_{\alpha_0\beta_0}^{\gamma_0},
\end{eqnarray}
\begin{eqnarray}
\bigl\{H_0,\,T_{\alpha_0}\bigr\} &= & T_{\beta_0}V_{\alpha_0}^{\beta_0},
\label{1}
\end{eqnarray}

\noindent where the structure functions $U_{\alpha_0\beta_0}^{\gamma_0}$
possess the properties of generalized antisymmetry
$U_{\alpha_0 \beta_0}^{\gamma_0} =  -(-1)^{\epsilon_{\alpha_0}
\epsilon_{\beta_0}}U_{\beta_0 \alpha_0}^{\gamma_0}$.

Let us introduce an extended phase space $\Gamma$ parametrized by the
following canonical variables:

\begin{eqnarray}
&&\Gamma = (P_A,Q^A) = (p_i,q^i; {\cal P}_{\alpha_0 a}, {\cal C}^{\alpha_0 a},
a=1,2; \cdot\cdot\cdot)\nonumber\\
&&\epsilon(P_A) = \epsilon(Q^A) =  \epsilon_A, \;\;
\epsilon({\cal P}_{\alpha_0 a})=\epsilon_{\alpha_0}+1,
\label{2}
\end{eqnarray}

\noindent where the dots stand for a set of possible additional auxiliary
canonical variables (see \cite{Lavrov1,Lavrov2}).
The explicit structure of the extended phase space (\ref{2}) depends on the
properties of the constraints $T_{\alpha_0} =  T_{\alpha_0}(p_{i},q^{i})$
of the initial dynamical system as well as on the existence of non-trivial
solutions to the generating equations of the formalism. Here, we are not
concerned in the explicit structure of $\Gamma$, as we will discuss only the
most general features of the canonical $osp(1,2)$ quantization.

The key role in the canonical scheme of $osp(1,2)$ quantization belongs
to the set of generating functions
$\Omega_{m}^a$, $\Omega_\alpha$ and $\cal H$, with $a =
  1,2$; \,$\alpha =  0,+,-$;
\,$\epsilon(\Omega_{m}^a) =  1$, $\epsilon(\Omega_\alpha) =  0$ and
 $\epsilon({\cal H}) = 0$.
The functions
$\Omega_m^a$ and $\Omega_\alpha$ satisfy the generating equations

\begin{eqnarray}
\bigl\{\Omega_{\alpha},\,\Omega_{\beta}\bigr\} &= & {\epsilon_{\alpha
\beta}}^\gamma \Omega_\gamma,
%\label{3}
\end{eqnarray}
\begin{eqnarray}
\bigl\{\Omega_{\alpha},\,\Omega_m^a\bigr\} &= & \Omega_m^b
{(\sigma_\alpha)_b}^a,
%\label{4}
\end{eqnarray}
\begin{eqnarray}
\bigl\{\Omega_m^a,\,\Omega_m^b\bigr\} &= &
-m^2(\sigma^\alpha)^{ab}\Omega_\alpha,
%\label{5}
\end{eqnarray}

\noindent where $m$ is a constant (mass) parameter,

\begin{eqnarray}
\label{sigma}
(\sigma^\alpha)^{ab}= \epsilon^{ac}(\sigma^\alpha)_c^{\,\,\,b}=
(\sigma^\alpha)^a_{\,\,\,c}\epsilon^{cb}=
\epsilon^{ac}(\sigma^\alpha)_{cd}\epsilon^{db},\,\,\,\,
(\sigma^\alpha)_a^{\,\,\,b}= -(\sigma^\alpha)^b_{\,\,\,a},
\end{eqnarray}

\noindent while $\sigma_\alpha$ generate the group of special linear
transformations with the $sl(2)$ algebra

\begin{eqnarray}
\label{algebra}
\sigma_\alpha\sigma_\beta & =  & g_{\alpha\beta}+{1\over 2}\epsilon_
{\alpha\beta\gamma}\sigma^\gamma, \,\,\, \sigma^\alpha=
g^{\alpha\beta}\sigma_\beta, \,\,\, Tr(\sigma_\alpha\sigma_\beta)=
2g_{\alpha\beta},\nonumber \\
g^{\alpha\beta} & =  & \left( \begin{array}{c}
1\,\,\,  0 \,\,\, 0\\0\,\, \, 0 \,\, \,2\\ 0\,\,\,  2 \,\,\, 0
\end{array} \right) , \,\,
g^{\alpha\gamma}g_{\gamma\beta}= \delta^\alpha_\beta,
\end{eqnarray}

\noindent where $\epsilon_{\alpha\beta\gamma}$ is the antisymmetric
tensor $\epsilon_{0+-}= 1$. The algebra of functions (8), (9),
(10) is isomorphic to the $osp(1,2)$-superalgebra \cite{Pais}. Here
it should be noted that the right-hand side of (10) for $m\neq 0$ is
a generalization of the conventional extended BRST relations of the
$Sp(2)$-formalism \cite{Lavrov1,Lavrov2,Lavrov3} (for earlier
discussions of the extended BRST symmetry in the generalized
canonical formalism, see also \cite{Hwang}).

The functions $\Omega_m^a$ and $\Omega_\alpha$ should be
considered as the generators of extended BRST and $Sp(2)$
transformations, respectively.  In its turn, the boson function $\cal
H$ satisfies the generating equations

\begin{eqnarray}
\bigl\{{\cal H},\,\Omega_\alpha\bigr\} &= & 0,
\label{5}
\end{eqnarray}
\begin{eqnarray}
\bigl\{{\cal H},\,\Omega_m^a\bigr\} &= & 0.
\label{6}
\end{eqnarray}

\noindent These equations can be interpreted as a requirement
of invariance of $\cal H$ under the transformations of the $Sp(2)$ and
extended BRST symmetries, respectively.

We shall now determine the unitarizing Hamiltonian $H$, in terms of $\cal H$
and $\Omega_m^a$, by the formula

\begin{equation}
H =  {\cal H} + \frac{1}{2} \epsilon_{ab}
\bigl\{\bigl\{\Phi,\,\Omega_m^b\bigr\},\,\Omega_m^a\bigr\} + m^2\Phi,
\label{7}
\end{equation}

\vspace{1cm}

\noindent where $\Phi$ is a boson function fixing a concrete choice of
admissible gauge. In what follows we will require $\Phi$ to be an
$Sp(2)$-scalar, i.e.

\begin{equation}
\bigl\{\Phi,\,\Omega_\alpha\bigr\} =  0.
\label{8}
\end{equation}

\noindent The extended BRST and Sp(2) transformations of canonical variables
are given, respectively, by

\begin{eqnarray}
\label{brst}
\delta\Gamma =  \bigl\{\Gamma,\,\Omega_m^a\bigr\}\mu_a;\,\,\epsilon
(\mu_a) =  1,\\
\label{9}
\delta\Gamma =
\bigl\{\Gamma,\,\Omega_\alpha\bigr\}\eta^\alpha;\,\,\,\epsilon
(\eta^\alpha) =  0.
\label{s}
\end{eqnarray}

\noindent Here, $\mu_a$ form an $Sp(2)$-doublet of constant
Grassmann parameters, and $\eta^\alpha$ are bosonic parameters.
Thus, an essential property of the Hamiltonian $H$ (\ref{7})
is its invariance

%\newpage

\begin{equation}
\delta H =  \bigl\{H,\,\Omega_m^a\bigr\}\mu_a =  0,\\
\label{10}
\end{equation}

\begin{equation}
\delta H =  \bigl\{H,\,\Omega_\alpha\bigr\}\eta^\alpha =  0.
\label{11}
\end{equation}

\noindent The invariance (\ref{10}) and (\ref{11}) follows from Eqs.
(\ref{5}) and (\ref{6}) as well as from the Jacobi identities for
$\Omega_m^a$ and $\Omega_\alpha$.

\section {Gauge dependence and Ward identities}

Let us define the vacuum functional $Z_\Phi$ in terms of the following
functional integral:

\begin{equation}
Z_\Phi =  \int D\Gamma \exp \bigg\{\frac{i}{\hbar}\int dt(P_A \dot{Q}^A -
H)\bigg\}.
\label{12}
\end{equation}

\vspace{1cm}

\noindent Making a change of the gauge $\Phi \rightarrow \Phi +
\Delta\Phi$
in (\ref{12}), we obtain

\begin{eqnarray}
Z_{\Phi+\Delta\Phi} &= & \int D\Gamma \exp \bigg\{
\frac{i}{\hbar}\int dt(P_A \dot{Q}^A - H \nonumber \\
&&- \frac{1}{2} \epsilon_{ab}
\bigl\{\bigl\{\Delta\Phi,\,\Omega_m^b\bigr\},\,\Omega_m^a\bigr\}
- m^2\Delta\Phi)\bigg\}.
\label{13}
\end{eqnarray}

\vspace{1cm}

\noindent The term $\frac{1}{2} \epsilon_{ab}
\bigl\{\bigl\{\Delta\Phi,\,\Omega_m^b\bigr\},\,\Omega_m^a\bigr\}$ in
(\ref{13}) can be compensated for by the change of integration variables
$\Gamma \longrightarrow \Gamma + \delta \Gamma$, where

\begin{equation}
\delta \Gamma =  \frac{i}{2\hbar}\bigl\{\Gamma,\,\Omega_m^a\bigr\}
\epsilon_{ab} \bigl\{\Omega_m^b,\,\Delta\Phi\bigr\}.
\label{14}
\end{equation}

\noindent On the other hand, it is impossible to cancel the term $m^2 \Delta
\Phi$ by using transformations of the form (\ref{s}) with any functions
$\eta^a$ due to the fact that matrices $\sigma_{\alpha}$ are traceless.

By comparison with (\ref{12}), we find that the term $ m^2\Delta\Phi$
violates the independence of the vacuum functional $Z_\Phi$ from the
choice of gauge. Hence,
$Z_{\Phi + \Delta\Phi} \neq Z_\Phi$, and therefore in the case $m \neq 0$
the $S-matrix$ within the formalism of canonical
$osp(1,2)$ quantization becomes gauge-dependent. Taking the limit
$m\rightarrow 0$, the gauge independence of the
$S-matrix$ and vacuum functional (\ref{12}) are restored. Moreover, in
this limit the vacuum functional (\ref{12}) is reduced to the
well-known answer of the generalized canonical $Sp(2)$ formalism
\cite{Lavrov1,Lavrov2}.

Finally, we shall derive the Ward identities. To begin with, we assume
that the $Sp(2)$-symmetries are realized on the variables
$\Gamma$ as rotations of $\Gamma$ with respect to the $Sp(2)$ index only

\begin{equation}
\{\Gamma,\,\Omega_a\} =  \Sigma_a\cdot\Gamma,
\label{15}
\end{equation}

\vspace{1cm}

\noindent where $\Sigma_a$ is a constant matrix.
In fact, this is a requirement that the transformations realized on
the canonical variables form a closed algebra. Let us also consider the
generating functional

\begin{eqnarray}
Z(J,\Gamma_{a}^*,\bar{\Gamma}) &= & \int D\Gamma \exp \bigg\{\frac{i}{\hbar}
\int dt(P_{A} \dot{Q}^A
- H + J\Gamma \nonumber \\
 &&+ \Gamma_{a}^\ast \{\Gamma,\,\Omega_m^a\} + \frac{1}{2}
\bar{\Gamma} \epsilon_{ab}
\bigl\{\bigl\{\Gamma,\,\Omega_m^b\bigr\},\,\Omega_m^a\bigr\})\bigg\}.
\label{16}
\end{eqnarray}

\vspace{1cm}

\noindent Given the above generating functional, the Green functions of the
theory with the Hamiltonian H are calculated through taking derivatives with
respect to the sources
J for $\Gamma_{a}^\ast =  \bar{\Gamma} =  J =  0$. In (\ref{16}) we
have introduced
additional sources $\Gamma_a^\ast$ to the transformation
$\{\Gamma,\,\Omega_m^a\}$
and a source $\bar{\Gamma}$ to the generator $\frac{1}{2}\epsilon_{ab}
\bigl\{\bigl\{\Gamma,\,\Omega_m^b\bigr\},\,\Omega_m^a\bigr\}\}$.
We shall now make the change of variables (17) in the functional
integral
(\ref{16}). Using the invariance property of $H$ (\ref{6}) as well as
the fact that the Berezinian of this change is equal to
$\exp \bigl\{-\int dt\bigl\{\Omega_m^a,\,\mu_a\bigr\}\bigr\} = 1$,
we obtain the following Ward identities for the generating functional
$Z$ in (\ref{16}):

\newpage

\begin{eqnarray}
J \, \frac{\delta Z}{\delta\Gamma_a^\ast}
- \epsilon^{ab} \, \Gamma_b^\ast \, \frac{\delta Z}{\delta\bar{\Gamma}}
+ \frac{1}{2}m^2(\sigma^\alpha)^{ab} \, \Gamma_b^\ast \, \Sigma_\alpha \, \frac{\delta Z}{\delta J}
+ \frac{1}{2}m^2 \, \bar{\Gamma} \, \frac{\delta Z}{\delta \Gamma_a^\ast}\nonumber\\
+ \frac{1}{2}m^2{(\sigma^\alpha)^a}_b \, \bar{\Gamma} \, \Sigma_\alpha
\, \frac{\delta Z}{\delta \Gamma_b^\ast} &= & 0.
\label{17}
\end{eqnarray}

\vspace{1cm}

\noindent In the functional integral (\ref{16}) we shall make the change of
variables (\ref{s}). With allowance for the invariance of $H$ and the fact
that the Berezinian of this change is equal to unity, we obtain the following
Ward identities for the generating functional Z in (\ref{16}):

\begin{eqnarray}
J \,\Sigma_\alpha \, \frac{\delta Z}{\delta J}
+ \Gamma_a^\ast \, \Sigma_\alpha \, \frac{\delta Z}{\delta \Gamma_a^\ast}
+ \bar{\Gamma}\, \Sigma_\alpha \, \frac{\delta Z}{\delta \bar{\Gamma}} =0.
\label{18}
\end{eqnarray}

\vspace{1cm}

\section{Conclusion}

In this paper we have considered a possibility to generalize the
canonical version of extended BRST quantization
\cite{Lavrov1,Lavrov2,Lavrov3}, i.e. by extending it to a canonical
$osp(1,2)$ quantization. This generalized canonical quantization can be
considered as a canonical counterpart to the $osp(1,2)$ - covariant
Lagrangian quantization \cite{Geyer}. As has been shown above, the
$osp(1,2)$ symmetry group permits introducing mass terms without
breaking the extended BRST symmetry.  Mass terms are important as a
means applied to solve the problem of infrared divergences.  Following
this consistent formulation of osp(1,2) approach, we have shown that
the vacuum functional and $S$-matrix with massive field terms are no
longer gauge-invariant.  Thus, it is evident that after performing the
renormalization procedure we need to take the massless limit in order
to get sensible physical answers because, in any case, physical results
do not depend on the gauge.

\section{Acknowledgments}
The work has been supported in part by FAPEMIG, Brazilian Research
Council. One of the authors (PML) appreciates partial support from\linebreak
FAPEMIG grant code number CEX-1308/97. The work of PML has also been
supported by the grant of Ministry of General and Professional
Education of Russian Federation in field of Basic Natural Sciences, as
well as by the grants INTAS 96-0308 and RFBR-DFG 96-02-00180.
The authors are grateful to the referee for critical remarks.


\begin{thebibliography} {99}

\bibitem{Lavrov1}
I.A. Batalin, P.M. Lavrov and I.V. Tyutin, J.Math.Phys.31 (1990) 6.

\bibitem{Lavrov2}
I.A. Batalin, P.M. Lavrov and I.V. Tyutin, J.Math.Phys.31 (1990) 2708.

\bibitem{Lavrov3}
I.A. Batalin, P.M. Lavrov and I.V. Tyutin, Int.J.Mod.Phys. 6 (1990) 3599.

\bibitem{Geyer}
B. Geyer, P.M. Lavrov and D. M\"ulsch, Acta Phys. Pol. B,
29 (1998) 2637.

\bibitem{Lavrov4}
I.A. Batalin, P.M. Lavrov and I.V. Tyutin, J.Math.Phys.31 (1990) 1487.

\bibitem{Lavrov5}
I.A. Batalin, P.M. Lavrov and I.V. Tyutin, J.Math.Phys.32 (1991) 532.

\bibitem{Lavrov6}
I.A. Batalin, P.M. Lavrov and I.V. Tyutin, J.Math.Phys.32 (1991) 2513.

\bibitem{Vilk}
E.S. Fradkin and G.A. Vilkovisky, Phys.Lett. B55, 224 (1975).

\bibitem{Bat}
I.A. Batalin and G.A. Vilkovisky, Phys.Lett. B69, 309 (1977).

\bibitem{Frad}
E.S. Fradkin and T.E. Fradkina, Phys.Lett. B72, 343 (1978).

\bibitem{Pais}
A. Pais and V. Rittenberg, J.Math.Phys.16 (1975) 2062;\\
W. Nahm and V. Rittenberg, J.Math.Phys.18 (1976) 146, 155;\\
F.A. Berezin and V.N. Tolstoy, Commmun. Math Phys.78 (1981) 409;\\
L.  Frappat, P. Sorba and A. Sciarrino, Dictionary on Lie
superalgebras, ENSLAPP-AL-600/96, hep-th/9607161.
\bibitem{Hwang} S.
Hwang, Nucl. Phys. B231, 386 (1984);\\
H. Aratyn, R. Ingermanson and A.J.  Niemi, Phys. Lett. B189, 427 (1987);
Nucl. Phys. B307, 157 (1988); \\
V.P.  Spiridonov, Nucl. Phys. B308, 527 (1988).

\end{thebibliography}
\end{document}